\documentclass[twocolumn,prl,groupedaddress]{revtex4}%
\usepackage{amssymb}
\usepackage[dvips]{graphicx}
\usepackage{dcolumn}
\usepackage{bm}
\usepackage{amsmath}
\usepackage{amsfonts}%
\newcommand{\OLD}[1]{{\tiny }}

\begin{document}
\title{ Non-equilibrium conductance of a three-terminal quantum dot in the Kondo regime:\\Perturbative Renormalization Group}
\author{N. Shah and A. Rosch}
\affiliation{Institut f\"{u}r Theoretische Physik, Universit\"{a}t zu K\"{o}ln, 50937
K\"{o}ln, Germany. }

\begin{abstract}
Motivated by recent experiments, we consider a single-electron transistor in
the Kondo regime which is coupled to three leads in the presence of large bias
voltages. Such a steady-state non-equilibrium system is to a large extent
governed by a decoherence rate $\Gamma$ induced by the current through the
dot. As the two-terminal conductance turns out to be rather insensitive to
$\Gamma$, we study the conductance in a three-terminal device using
perturbative renormalization group and calculate the characteristic splitting
of the Kondo resonance. The interplay between potential biases and anisotropy
in coupling to the three leads determines $\Gamma$ and the conditions for
strong coupling.\vspace*{-3mm}

\end{abstract}
\maketitle

With the advent of highly tunable strongly correlated systems it has not only
become possible to systematically realize and investigate well-studied
theoretical models but also to pose new theoretical questions of fundamental
nature. Though non-equilibrium physics in itself has a long history, a lot
remains to be unfolded in the context of strongly correlated quantum systems.
Since the single-impurity Kondo model, describing a spin coupled to conduction
electrons, is not only one of the best understood theoretical prototype but
also the one that has state-of-the-art experimental realization in tunable
nano-systems involving quantum dots \cite{cronenwett1998,goldhaber1998},
single molecules \cite{park2002,liang2002} and perhaps cold atoms in the near
future \cite{schlosser2001}, it is most natural and crucial to ask: what
happens at the juncture of Kondo and non-equilibrium physics?

In equilibrium, the Kondo coupling grows from \textquotedblleft
weak\textquotedblright\ to \textquotedblleft strong\textquotedblright\ on
lowering the temperature and finally the spin is screened. An applied bias
voltage $V$ influences the Kondo physics in at least two ways. First, it
prohibits resonant spin-scattering processes among the leads by splitting the
relevant Fermi energies by $V$. More interesting and less obvious
\cite{meir1993,kaminski2000,rosch2001,rosch2003,paaske2004} is the fate and
role of intra-lead resonant spin-flip couplings. Will they grow on reducing
the energy or will the problem remain weakly coupled for large voltages? At
least for symmetric coupling, the Kondo model remains at weak coupling for
$V\gg T_{K}$ ($T_{K}$ being the equilibrium Kondo temperature), as the large
current through the dot induces noise and decoheres the resonant spin-flip
processes as shown in Refs. \cite{rosch2001,rosch2003,paaske2004,rosch2005}.
The physics is then to a large extent controlled by the corresponding
decoherence rate $\Gamma$. Therefore the question arises which experimental
quantity can be used to investigate this second mechanism. Unfortunately, the
usual two-terminal conductance is almost completely blind to this kind of
physics as it it dominated by (non-resonant) inter-lead
processes\cite{magRemark}. However, a configuration where three leads are
coupled to a quantum dot has much more flexibility and is indeed sensitive to
$\Gamma$ as shown below.

Such systems have recently been realized by three
groups\cite{franceschi2002,leturcq2005,toonen2005}. Francheschi et al.
\cite{franceschi2002} used a clever indirect realization in a dot connected to
one lead and one mesoscopic wire driven out of equilibrium. The device of
Leturcq et al., in contrast, is based on a small ring with three leads
attached\cite{leturcq2005}. They managed to operate this device in the Kondo
regime, where effectively a single spin is localized (or rather delocalized)
on the ring. A magnetic field allows to fine-tune the wave function and
therefore to adjust the exchange couplings to the leads. A weakly coupled
third lead was proposed \cite{sun2001,lebanon2002,cho2003,sanchez2005} also as
a way to experimentally \textit{see }the splitting of the Kondo resonance by a
bias voltage, which lies beyond the scope of a two-terminal measurement
\textit{-- }this splitting has indeed been observed by these experiments.

In this letter, we carry out the non-equilibrium perturbative Renormalization
Group (RG) analysis \cite{rosch2005} of a three-terminal Kondoesque device and
calculate the conductance in the presence of applied voltage biases. We model
the system by the Hamiltonian
\begin{equation}
H=\sum_{\mathbf{\alpha k\sigma}}\varepsilon_{\mathbf{k}}c_{\alpha
\mathbf{k\sigma}}^{\dagger}c_{\alpha\mathbf{k\sigma}}+\frac{1}{3}\sum
_{\alpha\alpha^{\prime}\mathbf{kk}^{\prime}\sigma\sigma^{\prime}}%
J_{\alpha\alpha^{\prime}}S_{\lambda}c_{\alpha\mathbf{k\sigma}}^{\dagger}%
\sigma_{\sigma\sigma^{\prime}}^{\lambda}c_{\alpha^{\prime}\mathbf{k}^{\prime
}\mathbf{\sigma}^{\prime}} \label{H}%
\end{equation}
where $\alpha=1,2,3$ denote the three leads, each with a distribution function
$f_{\alpha}(\omega)=1/(1+\exp\beta(\omega-\mu_{\alpha}))$ with the
electrochemical potential $\mu_{\alpha}$ incorporating the applied voltages.
The key point is that in non-equilibrium, all electrons in the energy windows
corresponding to the applied voltages, contribute to the low energy properties
and the effective renormalized exchange coupling of the electrons from the
leads with the local \textquotedblleft spin\textquotedblright\ depends on
their energy. We derive the RG\ equations for these frequency-dependent
coupling functions (denoted by $g_{\alpha\alpha^{\prime}}(\omega)$) to leading
order in the small parameter $1/\ln[\max(\left\vert \mu_{\alpha}-\mu_{\beta
}\right\vert ,\left\vert \omega-\mu_{\alpha}\right\vert )/T_{K}]$. The
equations read
\begin{equation}
\frac{\partial g_{\alpha\alpha^{\prime}}(\omega)}{\partial\ln D}=-\frac{2}%
{3}\sum\limits_{\beta=1}^{3}g_{\alpha\beta}(\mu_{\beta})g_{\beta\alpha
^{\prime}}(\mu_{\beta})\Theta(D-\left\vert \omega-\mu_{\beta}\right\vert )
\label{RGeq}%
\end{equation}
with the initial conditions at the bare cutoff $D=D_{0}$ given by
$g_{\alpha\alpha^{\prime}}(\omega)=\rho J_{\alpha\alpha^{\prime}}$, where
$\rho$ is the density of states. The meaning is transparent: At each value of
the cut-off $D,$ an electron of lead $\beta$ at $\mu_{\beta}$ contributes to
the renormalization of $g_{\omega}(D)$ via a second order virtual process only
for frequencies that lie within the range of the cut-off from $\mu_{\beta}$.
By choosing $\omega=\mu_{\beta},$ the RG\ equations are first solved for all
the constants $g_{\alpha\beta}(\mu_{\beta})$ on the RHS to obtain their entire
$D-$dependence, and then in the second step $g_{\alpha\alpha^{\prime}}%
(\omega)$ is obtained by using in the RHS the pre-evaluated constants at every
$D$.

As we have emphasized, the crucial element in the non-equilibrium is the
so-called decoherence rate\cite{rosch2003}%
\begin{equation}
\Gamma=\frac{4\pi}{9\hbar}\sum\limits_{\alpha,\alpha^{\prime}}\int_{-\infty
}^{\infty}d\omega g_{\alpha\alpha^{\prime}}^{2}(\omega)f_{\alpha}%
(\omega)(1-f_{\alpha^{\prime}}(\omega))\label{gamma}%
\end{equation}
responsible for stopping the RG\ flow before the strong coupling regime given
by the above RG\ equation is reached. Implementation of this effect in the
RG\ procedure amounts to replacing the $\Theta$ function in Eq. (\ref{RGeq})
by $\Theta(D-\widetilde{\left\vert \omega-\mu_{\beta}\right\vert })$ (with the
short-hand $\widetilde{\omega}\equiv\sqrt{\omega^{2}+\Gamma^{2}}$) and
evaluating the value of $\Gamma$ self-consistently.

When the Hamiltonian is assumed to be derived from an underlying Anderson
model then the $J_{\alpha\alpha^{\prime}}$ are separable ($J_{\alpha
\alpha^{\prime}}\propto t_{\alpha}t_{\alpha^{\prime}}$ where $t_{\alpha}$ is
the hybridization) and only a single channel couples to the dot with
$T_{K}=D_{0}\exp[(3/(2\rho(J_{11}+J_{22}+J_{33}))]$ (to one-loop order). Since
the matrix structure of the general RG\ equations respects this property, we
can express $g_{\alpha\alpha^{\prime}}(\omega)=3g_{\omega}n_{\alpha}%
n_{\alpha^{\prime}}$ where $n_{\alpha}^{2}=J_{\alpha\alpha}/(J_{11}%
+J_{22}+J_{33})$ is the the lead anisotropy parameter (so that $n_{1}%
^{2}+n_{2}^{2}+n_{3}^{2}=1$) and the initial condition is given by $g_{\omega
}(D_{0})=1/[2\ln(D_{0}/T_{K})]$.

For concreteness, let us assume that $\mu_{31}>\mu_{32}>\mu_{21}$ with
$\mu_{\alpha\beta}\equiv\left\vert \mu_{\alpha}-\mu_{\beta}\right\vert $. When
solving the RG equations for $g_{\alpha\beta}(\mu_{\beta})$ as a function of
the running cutoff $D$, four different regimes have to be considered. First,
for $D>\mu_{31}$ one recovers the usual equilibrium RG equations which are
easily solved analytically. This is not possible in the second regime,
$\mu_{31}>D>\mu_{32}$. However, as $\ln[\mu_{31}/\mu_{32}]\leq\ln2\ll\ln
\mu_{31}/T_{K}$ (as $\mu_{32}>\mu_{21}$ by definition), the contributions from
this intermediate regime are small and formally subleading in $1/\ln[\mu
_{31}/T_{K}]$. We can therefore safely take our formulas for $D>\mu_{31}$ and
use them even for $D>\mu_{32}$. In the two remaining regimes, $\mu_{32}\geq
D\geq\mu_{21}$ and finally $\mu_{21}\geq D\geq0$ again an analytical solution
can be obtained. When calculating $g_{\alpha\beta}(\omega)$ in a second step
by simple integration of the RHS of Eq. (\ref{RGeq}), we do not use any
further approximation and treat all four regimes equally (thereby improving
also the results for $g_{\alpha\beta}(\mu_{\beta})$ for $\mu_{31}>D>\mu_{32}$
as discussed below). The approximation improves with decreasing $\mu_{21}$ and
for $\mu_{21}=0$ one recovers correctly the two-lead case for which the one
loop RG\ equations can be solved analytically.

\begin{figure}
[ptb]
\begin{center}
\includegraphics[
height=2.4716in,
width=3.2063in
]%
{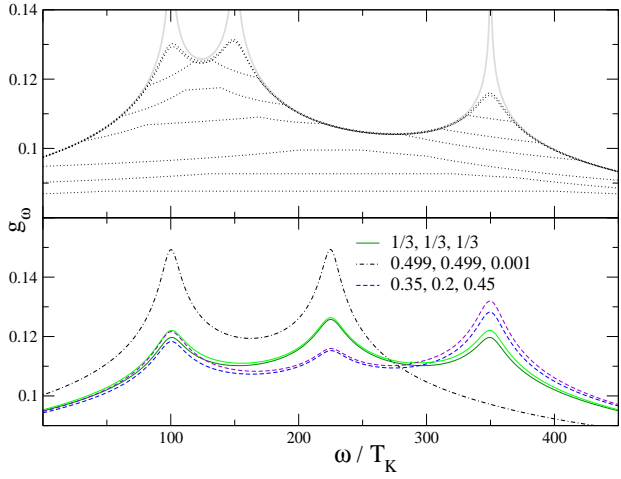}%
\caption{Upper panel shows the scaling of $g_{\omega}$ as the cut-off is
decreased down to zero from $D=300T_{K}$, for $\mu_{1}=100T_{K},$ $\mu
_{2}=150T_{K},$ $\mu_{3}=350T_{K}$ and $n_{1}=n_{2}=n_{3}$. The light grey
plot shows the $g_{\omega}$ obtained if $\Gamma$ were to be $T_{\ast}$. Lower
panel shows the renormalized $g_{\omega}$($D=0$) for $\mu_{1}=100T_{K},$
$\mu_{2}=225T_{K},$ $\mu_{3}=350T_{K}$ for different $n_{1},n_{2},n_{3}$ shown
in the legends. In the same line style as numerical plots, also shown in both
panels are the plots for $g_{\omega}$ obtained analytically (upper curves when
they do not coincide).\vspace*{-3mm}}%
\label{fig: RG}%
\end{center}
\end{figure}

The virtue of having an analytical solution is that it provides insight into
how the four independent dimensionless parameters (two $\mu_{\alpha\beta
}/T_{K}$ and two $n_{\alpha}^{2}$ ) enter the renormalized $g_{\omega}$ at
various stages of the scaling procedure and for what values the strong
coupling regime is approached. The role of energy scales is exemplified by the
sheer number of regions (already seven for $\mu_{1}<\omega<\mu_{3}$) into
which the entire frequency range gets divided, most having a different closed
form solution for $g_{\omega}$. For the frequency range $\mu_{3}\geq\omega
\geq\mu_{2}+\mu_{21}$, we obtain for $D\rightarrow0$%

\begin{align}
g_{\omega} &  =\frac{n_{1}^{2}}{2\ln\left[  \frac{\widetilde{\left\vert
\omega-\mu_{1}\right\vert }}{T_{K}}\right]  }-\left[  n_{1}^{2}+\frac
{n_{2}^{2}}{n_{1}^{2}+n_{2}^{2}}\right]  \frac{1}{2\ln\left[  \frac
{\widetilde{\mu_{32}}}{T_{K}}\right]  }\nonumber\\
&  +\frac{n_{2}^{2}}{n_{1}^{2}+n_{2}^{2}}\frac{1}{2(n_{1}^{2}+n_{2}^{2}%
)\ln\left[  \frac{\widetilde{\left\vert \omega-\mu_{2}\right\vert }}{T_{K}%
}\right]  +2n_{3}^{2}\ln\left[  \frac{\widetilde{\mu_{32}}}{T_{K}}\right]
}\nonumber\\
&  +\frac{1}{2n_{3}^{2}\ln\left[  \frac{\widetilde{\left\vert \omega-\mu
_{3}\right\vert }}{T_{K}}\right]  +2(n_{1}^{2}+n_{2}^{2})\ln\left[
\frac{\widetilde{\mu_{32}}}{T_{K}}\right]  }\label{ga}%
\end{align}
while for $\mu_{2}\geq\omega\geq\mu_{1}$ we have
\begin{align}
g_{\omega} &  =\frac{1}{2n_{1}^{2}\ln\left[  \frac{\widetilde{\left\vert
\omega-\mu_{1}\right\vert }}{T_{K}}\right]  +2n_{2}^{2}\ln\left[
\frac{\widetilde{\mu_{21}}}{T_{K}}\right]  +2n_{3}^{2}\ln\left[
\frac{\widetilde{\mu_{32}}}{T_{K}}\right]  }\nonumber\\
&  +\frac{1}{2n_{2}^{2}\ln\left[  \frac{\widetilde{\left\vert \omega-\mu
_{2}\right\vert }}{T_{K}}\right]  +2n_{1}^{2}\ln\left[  \frac{\widetilde
{\mu_{21}}}{T_{K}}\right]  +2n_{3}^{2}\ln\left[  \frac{\widetilde{\mu_{32}}%
}{T_{K}}\right]  }\nonumber\\
&  +\frac{n_{3}^{2}}{2\ln\left[  \frac{\widetilde{\left\vert \omega-\mu
_{3}\right\vert }}{T_{K}}\right]  }-\frac{n_{3}^{2}}{2\ln\left[
\frac{\widetilde{\mu_{32}}}{T_{K}}\right]  }\nonumber\\
&  -\frac{1}{2(n_{1}^{2}+n_{2}^{2})\ln\left[  \frac{\widetilde{\mu_{21}}%
}{T_{K}}\right]  +2n_{3}^{2}\ln\left[  \frac{\widetilde{\mu_{32}}}{T_{K}%
}\right]  }.\label{gb}%
\end{align}
As a check for our approximative treatment of the RG equations, we compare
e.g. $g_{\mu_{3}}$ evaluated by taking $\omega=\mu_{3}$ in Eq. (\ref{ga}),
with its initial evaluation described above (not shown). The difference
$n_{1}^{2}(1/\ln\left[  \widetilde{\mu_{32}}/T_{K}\right]  -1/\ln\left[
\widetilde{\mu_{31}}/T_{K}\right]  )/2$ is even for the worst case ($\mu
_{32}=\mu_{31}/2$) only of order $\ln2/\ln^{2}\left[  \mu_{31}/T_{K}\right]  $
as expected. As Fig. \ref{fig: RG} shows, there is only a very slight
overestimation of $g_{\omega}$ as compared to the straightforward numerical
solution of Eq. (\ref{RGeq}) in this case. In all other cases the numerical
and analytical curves coincide.

It is important to check under what conditions the problem enters the strong
coupling regime where our approach is not valid any more. We can investigate
this question analytically by determining the scale $T_{\ast}^{\alpha}$ at
which $g_{\mu_{\alpha}}$diverges \textit{in the absence of }$\Gamma$.
Analyzing Eqs. (\ref{ga}),(\ref{gb}) , for arbitrary $\mu_{\alpha\beta
}>T_{\ast}$ we obtain
\begin{equation}
T_{\ast}=\max_{\alpha=1,2,3}[T_{\ast}^{\alpha}]\text{ \quad with \quad}%
\frac{T_{\ast}^{\alpha}}{T_{K}}=\prod_{\beta\neq\alpha}\left[  \frac{T_{K}%
}{\mu_{\beta\alpha}}\right]  ^{\left(  \frac{n_{\beta}}{n_{\alpha}}\right)
^{2}}\label{Tstar}%
\end{equation}
while if two of the three leads have the same electrochemical potential (more
precisely $\mu_{12}<T_{\ast}$, say) then%
\begin{equation}
T_{\ast}=\left[  \frac{T_{K}}{\mu_{32}}\right]  ^{\min\left[  (n_{2}^{2}%
+n_{1}^{2})/n_{3}^{2},\text{ }n_{3}^{2}/(n_{1}^{2}+n_{2}^{2})\right]
}.\label{Tstareq}%
\end{equation}
For $\Gamma=T_{\ast}$ at least one of the coupling constants diverges
$\propto1/\ln[\Gamma/T_{\ast}]$ (this is how we determine $T_{\ast}^{\alpha}$;
note that $\widetilde{x}\equiv\sqrt{x^{2}+\Gamma^{2}}$ in the terms
$\ln[\widetilde{x}/T_{K}]$). The perturbative RG\ is therefore controlled by
the ratio $\Gamma/T_{\ast}$. The larger is the ratio as compared to one, the
further away we are from the strong coupling coupling regime, and the more
accurate is our analysis.%

The upper panel of Fig. \ref{fig: RG} is a visual display of the scaling
process: $g_{\omega}$ for a generic case is plotted as the cut-off $D\ $is
gradually decreased; one can see the role of energy scales $\mu_{21}$,
$\mu_{31}$ and $D$, all measured in units of $T_{K}$. To clarify the role of
$\Gamma$, we also plot (in light grey) the $g_{\omega}$ obtained if $\Gamma$
were to be $T_{\ast}$. In the lower panel we illustrate how the interplay
between lead anisotropies and potential differences determines the
renormalization of $g_{\omega}$. For example, when one lead $\alpha$ is very
weakly coupled, there is no feature at $\omega=\mu_{a}$.

The physical quantity of interest is the non-equilibrium conductance matrix,
which, within the precision of our approach, can be evaluated by plugging the
renormalized coupling constants in the second order perturbative formula to
obtain
\begin{align}
G_{\alpha\beta} &  \equiv e\partial I_{\alpha}/\partial\mu_{\beta}=\frac
{\pi^{2}G^{0}}{3}\sum\limits_{\alpha^{\prime}\neq\alpha}[(\delta_{\alpha\beta
}-\delta_{\alpha^{\prime}\beta})g_{\alpha\alpha^{\prime}}^{2}(\mu_{\beta
})\nonumber\\
&  +\int_{\mu_{\alpha^{\prime}}}^{\mu_{\alpha}}d\varepsilon\frac{\partial
}{\partial\mu_{\beta}}g_{\alpha\alpha^{\prime}}^{2}(\varepsilon)]\label{G}%
\end{align}
where we have restricted to zero temperature and introduced the quantum of
conductance, $G^{0}=e^{2}/\pi\hbar$. Let us first consider the case when lead
one is a \textquotedblleft probing lead\textquotedblright\ i.e. it is very
weakly coupled to the dot ($n_{1}^{2}\ll1$). Then $G_{11}(\mu_{1})$ is simply
given by the first term of Eq. (\ref{G}) (the second term being negligible
owing to very weak dependence of $g^{2}(\varepsilon)$ on $\mu_{1}$) and traces
the two-lead spectral function of the quantum dot; we have explicitly
confirmed this equivalence. If we assume an underlying Anderson model as
above, we obtain $G_{11}(\mu_{1})/G^{0}=3\pi^{2}n_{1}^{2}(1-n_{1}^{2}%
)g_{\mu_{1}}^{2}$ with $g_{\mu_{1}}$ given by Eq. (\ref{gb}). As $\mu_{1}$ is
swept, $G_{11}$ peaks up at $\mu_{2}$ and at $\mu_{3}$ with the peak height at
$\mu_{2}$ given by%
\begin{equation}
G_{11}(\mu_{1}=\mu_{2})=\frac{3\pi^{2}n_{1}^{2}(1-n_{1}^{2})G^{0}}{\left(
2(1-n_{3}^{2})\ln\left[  \frac{\Gamma}{T_{K}}\right]  +2n_{3}^{2}\ln\left[
\frac{\widetilde{\mu_{32}}}{T_{K}}\right]  \right)  ^{2}}\label{Gpeak}%
\end{equation}
while that at $\mu_{3}$ obtained by interchanging $n_{2}\leftrightarrow n_{3}$
in the above expression. For $n_{1}^{2}\ll1$ the formula for $T_{\ast}$ (Eq.
(\ref{Tstar})) reduces to Eq. (\ref{Tstareq}) which now holds not only for
$\mu_{1}=\mu_{2}$ but for all values of $\mu_{1}$.%

\begin{figure}
[ptb]
\begin{center}
\includegraphics[
height=2.1378in,
width=3.208in
]%
{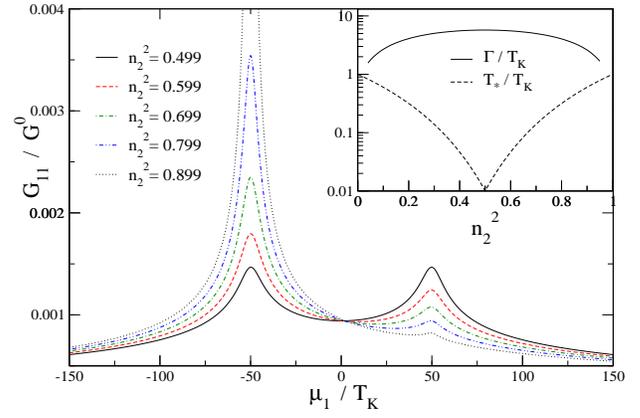}%
\caption{{}Conductance $G_{11}$ in the \textquotedblleft two-lead
spectroscopy\textquotedblright\ case ($n_{1}^{2}=0.002$)\ for $\mu
_{32}=100T_{K}$ and different values of $n_{2}^{2}$ and $n_{3}^{2}$. Inset
shows $\Gamma$ and $T_{\ast}$ on a log scale, as a function of $n_{2}^{2}$.}%
\label{fig: spec_aniso}%
\end{center}
\end{figure}

%

\begin{figure}
[ptb]
\begin{center}
\includegraphics[
height=2.166in,
width=3.237in
]%
{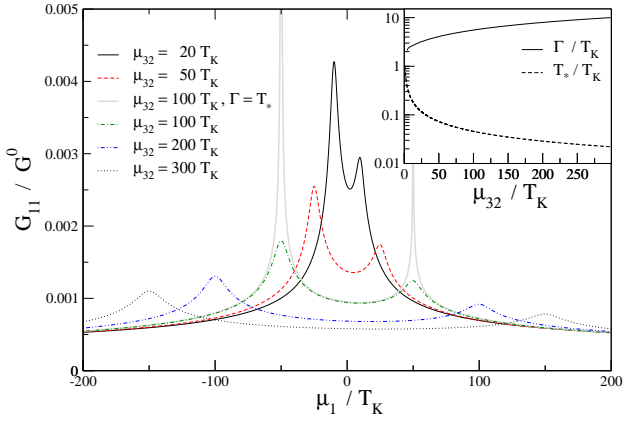}%
\caption{Conductance $G_{11}$ in the \textquotedblleft two-lead
spectroscopy\textquotedblright\ case ($n_{1}^{2}=0.002$) for $n_{2}%
=0.599,n_{3}=0.399$ and different values of $\mu_{32}$. The light curve shows
$G_{11}$ if $\Gamma$ were to be $T_{\ast}$. The inset shows $\Gamma$ and
$T_{\ast}$ on a log scale, as a function $\mu_{32}$.\vspace*{-1mm} }%
\label{fig: spec_V}%
\end{center}
\end{figure}

In Fig. \ref{fig: spec_aniso} we plot the conductance for different values of
the relative anisotropy between couplings to leads two and three, for a given
value of potential difference $\mu_{32}$, to emphasize the role of anisotropy.
The current from lead two to three is largest for $n_{2}\sim n_{3}=0.5$ and
accordingly $\Gamma$ (same for all $\mu_{1}$ as $n_{1}^{2}\ll1$) is maximum at
this point . On the other hand, $T_{\ast}$ is minimum at this point and the
ratio of $\Gamma/T_{\ast}$ largest (see inset). In the extreme anisotropy case
when only a single lead $\alpha$ is strongly coupled to the dot, one will
measure the equilibrium spectral function ($g(\omega)\ $is strongly
renormalized only at that $\mu_{\alpha}$). In Fig. \ref{fig: spec_V} we show
how the conductance changes as the potential difference $\mu_{32}$ is
increased, for a given relative anisotropy. To clarify how $\Gamma$ determines
the rounding of the peaks, the light line shows $G_{11}$ if $\Gamma$ were to
be $T_{\ast}$ (for $\mu_{32}=100T_{K}$). With increasing $\mu_{32}$, the value
of $\Gamma$ gets larger (since the current increases ) and $T_{\ast}$ gets
smaller, making the ratio $\Gamma/T_{\ast}$ increasingly larger (see inset).

\begin{figure}
[ptb]
\begin{center}
\includegraphics[
height=2.2557in,
width=3.237in
]%
{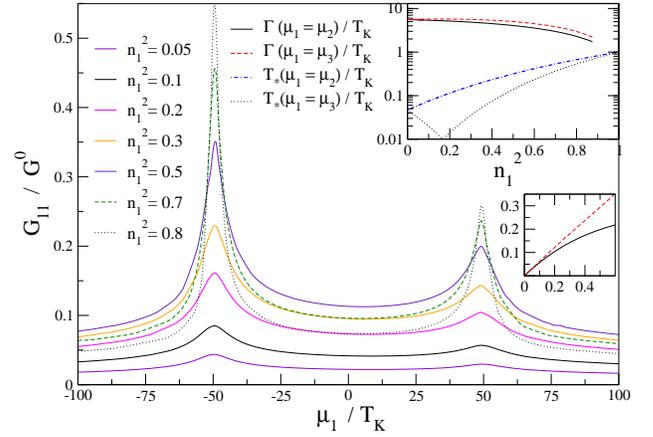}%
\caption{Plots of $G_{11}$ for increasing values of $n_{1}^{2}$ (ordered by
increasing peak heights) for fixed $n_{2}^{2}/n_{3}^{2}=1.5$ and $\mu
_{32}=100T_{K}$ . The corner inset shows $\Gamma$ and $T_{\ast}$ on a log
scale, as a function of $n_{1}^{2}$ for $\mu_{1}=\mu_{2}$, $\mu_{3}$. Other
inset shows the peak height $G_{11}(\mu_{1}=\mu_{3})$ as well as the part
coming only from the first term of Eq. \ref{G}. (dashed line)\ plotted against
$n_{1}^{2}$ .\vspace*{-3mm}}%
\label{fig: G}%
\end{center}
\end{figure}

To study other cases, we phrase the question: how does the conductance change
as we increase the coupling to lead one? We hold the relative anisotropy
between leads two and three to be constant as we increase the value of $n_{1}%
$. As can be seen from Fig. \ref{fig: G}, as we start increasing $n_{1}$, the
overall conductance $G_{11}$ begins to grow. Further, the second term in Eq.
(\ref{G}) is not negligible as in the \textquotedblleft two-lead
spectroscopy\textquotedblright\ case and makes a negative contribution; the
inset plots the peak height $G_{11}(\mu_{1}=\mu_{3})$ as well as the
contribution it gets from the first term to make the increasing deviation
between the two evident. As we keep increasing $n_{1}$, we start approaching
another limit in which lead one is strongly coupled and leads two and three
weakly coupled to the quantum dot: the peaks continue to grow, however, the
overall conductance for other values of $\mu_{1}$ begins to drop down. For the
sake of presentation we have chosen to focus on $G_{11}$ although it should be
clear that analogous arguments hold for the remaining elements of the
conductance matrix.%

As can be seen from the insets of Figs. \ref{fig: spec_aniso}%
,\ref{fig: spec_V},\ref{fig: G}, the ratio $\Gamma/T_{\ast}$ can become quite
large for some parameter range making our perturbative RG\ treatment extremely
accurate. Further more, only when the system approaches equilibrium and
$T_{\ast}\rightarrow T_{K},$ do we find that $\Gamma\rightarrow T_{\ast}$ and
the strong coupling regime is approached. This happens either when $\mu
_{32}\leq T_{K}$ for the $n_{1}^{2}\ll1$ case and all $\mu_{\alpha\beta}\leq
T_{K}$ for the general case, or in the extreme anisotropy limit when only one
of the three leads is strongly coupled to the dot irrespective of the values
of $\mu_{\alpha\beta}$. Our finding is in accordance with the expectation for
a Kondo model derived from an Anderson model: a strong coupling fixed point
exists only in equilibrium\cite{rosch2001}.

Possible extensions to this work include studying the case of a generalized
Kondo model with no underlying Anderson model and considering the effect of
excited levels or of a Zeeman field ($B\gg T_{K}$). For the specific set-up of
a quantum ring geometry enclosing a magnetic flux \cite{leturcq2005} it might
be interesting to investigate the chiral component to the
current\cite{chamon2003} in the Kondo regime and in non-equilibrium.

\begin{acknowledgments}
We thank K. Ensslin, R. Leturcq, J. Paaske and P. W\"{o}lfle for discussions
and the DFG for financial support.\vspace*{-4mm}
\end{acknowledgments}

\bibliographystyle{phrepetal}
\vspace*{-4mm}
\bibliography{strings,kondo2005,NonEq2005}

\begin{thebibliography}{10}

\bibitem{cronenwett1998}
S.~M. Cronenwett, T.~H. Oosterkamp, and L.~P. Kouwenhoven,
\newblock Science {\bf 281}, 540 (1998).

\bibitem{goldhaber1998}
D.~Goldhaber-Gordon {\it et~al.},
\newblock Nature {\bf 391}, 156 (1998).

\bibitem{park2002}
J.~Park {\it et~al.},
\newblock Nature {\bf 417}, 722 (2002).

\bibitem{liang2002}
W.~Liang {\it et~al.},
\newblock Nature {\bf 417}, 725 (2002).

\bibitem{schlosser2001}
N.~Schlosser {\it et~al.},
\newblock Nature {\bf 411}, 1024 (2001).

\bibitem{meir1993}
Y.~Meir, N.~S. Wingreen, and P.~A. Lee,
\newblock Phys. Rev. Lett. {\bf 70}, 2601 (1993).

\bibitem{kaminski2000}
A.~Kaminski, Y.~V. Nazarov, and L.~I. Glazman,
\newblock Phys. Rev. B {\bf 62}, 8154 (2000).

\bibitem{rosch2001}
A.~Rosch, J.~Kroha, and P.~W\"olfle,
\newblock Phys. Rev. Lett. {\bf 87}, 156802 (2001).

\bibitem{rosch2003}
A.~Rosch {\it et~al.},
\newblock Phys. Rev. Lett. {\bf 90}, 076804 (2003).

\bibitem{paaske2004}
J.~Paaske {\it et~al.},
\newblock Phys. Rev. B {\bf 70}, 155301 (2004).

\bibitem{rosch2005}
A.~Rosch {\it et~al.},
\newblock J. Phys. Soc. Jpn. {\bf 74}, 118 (2005).

\bibitem{magRemark}
Resonant inter-lead processes possible in the presence of a Zeeman field $B
  \sim V$ do not contribute to the current.

\bibitem{franceschi2002}
S.~De Franceschi {\it et~al.},
\newblock Phys. Rev. Lett. {\bf 89}, 156801 (2002).

\bibitem{leturcq2005}
R.~Leturcq {\it et~al.},
\newblock Phys. Rev. Lett. {\bf 95}, 126603 (2005).

\bibitem{toonen2005}
R.~C. Toonen {\it et~al.}, 2005,
\newblock private communication.

\bibitem{sun2001}
Q.~F.~Sun and H.~Guo,
\newblock Phys. Rev. B {\bf 64}, 153306 (2001).

\bibitem{lebanon2002}
E.~Lebanon and A.~Schiller,
\newblock Phys. Rev. B {\bf 65}, 035308 (2002).

\bibitem{cho2003}
S.~Y. Cho {\it et~al.},
\newblock Phys. Rev. B {\bf 68}, 125327 (2003).

\bibitem{sanchez2005}
D.~Sanchez and R.~L\'opez,
\newblock Phys. Rev. B {\bf 71}, 035315 (2005).

\bibitem{chamon2003}
C.~Chamon, M.~Oshikawa, and I.~Affleck,
\newblock Phys. Rev. Lett. {\bf 91}, 206403 (2003).

\end{thebibliography}

\end{document}